# FAULT BASED TECHNIQUES FOR TESTING BOOLEAN EXPRESSIONS: A SURVEY

[1]Usha Badhera
[2]Purohit G.N
[3]S.Taruna

Computer Science Department, Banasthali University, India

[1]ushas133@yahoo.com,
[2]gn_purohitjaipur@yahoo.co.in
[3,]staruna71@yahoo.com

## ABSTRACT

Boolean expressions are major focus of specifications and they are very much prone to introduction of faults, this survey presents various fault based testing techniques. It identifies that the techniques differ in their fault detection capabilities and generation of test suite. The various techniques like Cause effect graph, meaningful impact strategy, Branch Operator Strategy (BOR), BOR+MI, MUMCUT, Modified Condition/ Decision Coverage (MCDC) has been considered. This survey describes the basic algorithms and fault categories used by these strategies for evaluating their performance. Finally, it contains short summaries of the papers that use Boolean expressions used to specify the requirements for detecting faults. These techniques have been empirically evaluated by various researchers on a simplified safety related real time control system.

## KEYWORDS: -

Boolean expressions, Branch Operator Strategy (BOR), Meaningful Impact (MI), BOR+MI, Modified Condition/ Decision Coverage (MCDC), MUMCUT, fault detection.

## 1 INTRODUCTION

Software size and complexity is increasing that has made software testing a challenging exercise. The objective of testing is to determine error, which requires dynamic execution of test cases that consumes significant amount of time so it is important to investigate ways of increasing the efficiency and effectiveness of test cases.

Test case designing is one of the important factors that influence cost and coverage of testing. The cost depends on size of test suit and coverage on fault detection capabilities. Much research has been aimed at achieving high efficacy and reduced cost of testing by selecting appropriate test cases. Boolean expressions can be used to specify the requirements of safety-critical software like avionics, medical and other control software. These expressions can describe certain conditions of specifications, to model predicates and logical expressions. Test cases are generated on Boolean expressions which are capable of revealing faults in programs that are developed based on such specifications.

Many testing techniques have been proposed by various researchers to select test cases based on Boolean specifications; moreover test case generated by these methodologies can guarantee to detect certain type of faults. A literature search has revealed different Boolean specification testing techniques described through various research papers published from 1973 to 2011. This survey aims at presenting such techniques at one place and form a basis for comparison among these techniques.

| S.no | Authors | Method |
|------|---------|--------|
| 1. | W. R. Elmendrof | Cause effect graph |
| 2. | Myers | Algorithm CEG_Myers |
| 3. | Weuker et al. | Basic Meaningful Impact Strategy |
| 4. | Tai | Boolean Operator Strategy |
| 5. | Chilenski and Miller | MC/DC Coverage to software testing |
| 6. | Chen et al. | MUMCUT |

Table1: A chronological overview of various Fault Based testing techniques.

Boolean expressions are found in logical predicates inside programs and specifications which model complex conditions. Boolean predicate p with n variables requires 2n test cases in order to distinguish from any other predicate not equivalent to p. In practice, n can be quite large, there are examples of Boolean expressions with 30 or more conditions in an electronic flight implementation system, thus even for a rigid and simple formal specification exhaustive testing is not feasible as it becomes very expensive. In this paper, various approaches has been surveyed in which test cases are generated from Boolean expressions that target specific fault classes and test suites is reduced with respect to exhaustive testing. In this article, it is assumed that readers are familiar with notations and terminologies of Boolean expressions.

## 2. BOOLEAN SPECIFICATIONS BASED TESTING STRATEGIES- METHODS

A formal specifications of traffic collision avoidance system, TCAS II [7] uses AND/OR table, representation of Boolean expressions, to describe it. Logical expressions such as predicates in program source code modelled as Boolean expressions has been discussed in [6, 23],various methodologies have been proposed to select test cases based on Boolean expressions. Test cases generated by these methodologies guarantee to detect certain faults.

### 2.1 Experimental steps in empirical analysis of various testing techniques based on Boolean expressions

1. Boolean specification are selected and converted to Boolean expressions
2. For evaluating the performance of various techniques fault based approach is used. All faulty decisions are generated by mutation.
3. The test cases generated by specific strategy distinguish between original Boolean expression and the faulty one.
4. Effectiveness of test set is analysed by running test cases on the mutated Boolean expression and identifying what type of fault is captured.

### 2.2 Boolean specification testing techniques

### 2.2.1 Cause effect graphing

The cause effect graph was developed for system specification and test generation [4, 18]. It focuses on modelling dependency relationships among program input conditions known as causes, and output conditions known as, effects. The relationship is expressed visually in terms of cause-effect graph. The graph is a visual representation of logical relationship among inputs and outputs that can be expressed as a Boolean expression. One approach to test generation was to consider all possible combinations of causes of the CEG, which is exhaustive in nature but impractical as the test cases generated are exponential function of number of causes in the CEG. A practical test generation algorithm for CEGs was described by [18] which is referred to as algorithm CEG_Myers. Myers approach strengths and weaknesses has been investigated in[21] Myers process of creating decision table is inconsistent and ambiguous, other researchers [30, 31] has given algorithm for creating decision table from cause effect graph for generation of test cases.

**Algorithm for test generation by CEG_Myers**
The nodes *N* in graph are visited from effect to cause nodes.
1. If N is an OR node with "true" output value all combinations of inputs leading to a "t" output and having only one input being "t" are selected.
2. If *N* is an OR node with "false" output value all combinations of inputs leading to a "f" output are selected.
3. If *N* is an AND node with "true" output value all combinations of inputs leading to a "t" output are selected.
4. If *N* is an AND node with "false' output value all combinations of inputs leading to a "f" output are selected. However,
    - For the combination of all inputs being "f" only one test is selected for N and
    - For any combination with at least one input being "f" only one test is selected for each input being "t"

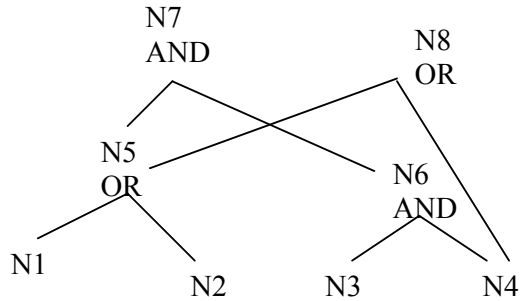

Figure1: A cause Effect Graph

Seven test cases selected for N7 by applying CEG_Myers approach{(t,f,t,t)(f,t,t,t)(f,f,t,t)(t,f,t,f)(t,f,f,t)(t,f,f,f)(f,f,f,f)} on figure1

**2.2.3 Boolean Operator testing Strategy**

BOR is a technique suitable for test generations for singular Boolean expression. It guarantees the detection of Boolean operator faults, including incorrect AND/OR operators and missing or extra Not operators.[23,24,25,26,27] showed that a BOR test set for a Boolean expression is effective in detecting various types of Boolean expression faults, including Boolean operator faults, incorrect Boolean variables and parentheses and their combinations.
However BOR strategy is not suitable for non singular expressions. When two tests are merged into one, if they contain conflicting values for the same variable, then the merge operation does not produce a test. This situation reduces the number of test cases generated, but also reduces fault-detection capability.
BOR strategy has excellent results when used with singular expressions, but needs to be modified when used with non singular expressions.

**Algorithm for test generation by BOR**
A test set S (E) is said to be a BOR test set for E if S (E) satisfies the BOR testing strategy for E. If E is a simple Boolean expression then the minimum BOR test set for E is given by {(t),(f)}. If E is a compound Boolean expression, then E can be represented as E1 op E2, where op could be either. or +, and E1, E2 are either simple or compound Boolean expressions. The following three rules show how to generate a BOR test set for E recursively. Assume that S (E1) and S (E2) are minimum BOR test sets for E1 and E2 respectively.

1. If E = E1.E2; then a minimum BOR set
S (E) is constructed as follows:
$S_t(E) = S_t(E1) \% S_t(E2)$
$S_f(E) = (S_f(E1) \times \{t_{E2}\}) \cup (\{t_{E1}\} \times S_f(E2))$
Where $t_{E1} \in St(E1)$, $t_{E2} \in S_t(E2)$, and
$(t_{E1}, t_{E2}) \in S_t(E)$
2. If E = E1+E2; then a minimum BOR set
S (E) is constructed as follows:
$S_f(E) = S_f(E1) \% S_f(E2)$
$S_t(E) = (S_t(E1) \times \{f_{E2}\}) \cup (\{f_{E1}\} \times S_t(E2))$
Where $f_1 \in Sf(E1)$, $f_{E2} \in S_f(E2)$, and
$(f_{E1}; f_{E2}) \in S_f(E)$
3. If E = ¬E1, then a minimum BOR set S (E)
Is constructed as follows:
$S_f(E) = S_t(E1)$
$S_t(E) = S_f(E1)$:

**Example:** Two minimum test sets generated for node N7 of figure 1 by applying BOR strategy
$S_t(N7) = \{(t,f,t,t)(f,t,t,t)\}$
$S_f(N7) = \{(f,f,t,t)(t,f,t,f)$ or $\{f,f,t,t)(f,t,t,f)(f,t,f,t)\}$

### 2.2.4 Basic Meaningful Impact (MI) testing strategy

MI for Boolean expressions was reported in [34]. It can be applied to singular or non-singular expressions. The strategy is based on detection of missing and/or extra negation operators on individual variables. The author reported good detection rates for different types of faults but the test case generation methodology requires that the Boolean expressions be in DNF Disjunctive Normal Form.

Once the expression is in the required format, the strategy first generates, for each term in the DNF, test cases that make the term true. (That makes the whole expression true) The test set for each term then contain only those test case that make other terms in the DNF false. In the second step of the strategy, each variable in each term is negated one at a time and test cases that make only this modified term true are considered. This set represents the test cases that make the original term false. But some of these test cases might still make the overall expression true because of other unmodified terms from the expression, such test cases are removed, and only those test cases that make the overall expression false are retained. This procedure is carried out for each variable in each term.

In the MI-Basic strategy for a Boolean expression, one test case from the set of unique true test cases for each term in the DNF is chosen to be part of St (E) for the expression. For the Sf (E), one test case from a false set of test cases for each term is selected. But Sf (E) may contain test cases that are in the false set for two or more terms. In the MI-MIN strategy, the test cases for making up St (E) are chosen as in the basic strategy, but a minimum set of test cases that satisfy the meaning impact strategy for the false outcome is chosen for Sf (E).

The results reported [34] of an empirical study done on twenty specifications written as Boolean expressions. The number of test cases generated using the MI-MIN strategy is a fraction of the exhaustive test cases that would be required. But the paper did not report the worst case size bounds in terms of number of operators. Fault detection rates for various fault types were also reported. The results showed good results which are comparable to those obtained using Foster's strategy, but with fewer test cases. Even though the strategy focuses on missing and extra NOT operators, it cannot guarantee detection of all such faults in the original expression. This is because the strategy works with DNF representation of the expression. Also, the study reported results only on single faults. Another problem with the MI-MIN strategy is that it sometimes generates extra test cases.

**Algorithm for Basic Meaningful Impact (MI) testing strategy**

A Boolean expression $E=e1+e2+...en$ in minimal DNF containing $n$ terms. Terms $e_i$, $1<=i<=n$ contains $l_j$ literals.

1. For each term $e_i$, $1<=i<=n$, construct $Te_i$ as the set of constraints that make $e_i$ true.
2. Let $TSe_i=Te_i-U^n_{j=1,i\neq j} Tej$. For $i \neq j$, $TSe_i \cap TSe_j=\varphi$
3. Construct $S^t_E$ by including one constraint from each $TSe_i, 1<=i<=n$
4. Let $e^j_i$ denote complimented term obtained by complementing $j^{th}$ literal in term $e_i$, for $1<=i<=n$ and $1<=j<=l_j$. Construct $Fe^j_i$ as the set of constraints that make $e^j_i$ true.
5. Let FS $e^j_i = Fe^j_i - U^n_{k=1} Te_{kz}$
6. Construct $S^f_E$ that is minimal and covers each $FS\ e^j_i$ at least once
7. Construct the desired constraint set for E as $S_E = S^t_E \cup S^f_E$

**Example**

Let $E=a(bc+\neg bd)$ the generated test cases by MI are
$S^t_E = \{(t,t,t,f)(t,f,f,t)\}$
$S^f_E = \{(f,t,t,f)(t,f,t,f)(t,t,f,t)(f,f,t,t)\}$

## 2.2.5 BOR+MI

The technique [20, 21], combines the BOR and MI. This hybrid algorithm partitions an input Boolean expression in to components such that BOR strategy can be applied to some and MI strategy to remaining components. The test constraints for individual components are combined using BOR strategy. Analytical and empirical results indicate that the BOR+MI algorithm usually produces a smaller test constraint set for Boolean expression then does the MI strategy. The BOR+MI strategy and MI strategy have comparable fault detection capability.

**Algorithm for BOR+MI**

1. Partition Boolean expression E into mutually singular components.
2. Generate test cases using BOR for each singular component.
3. Generate test cases using MI for non singular components.
4. Combine the constraints generated above.

**Example**

Let $E=a(bc+\neg bd)$ the generated test cases by BOR+MI are
$S^t_E = \{(t,t,t,f)(t,f,t,t)$
$S^f_E = \{(f,t,t,f)(t,f,t,f)(t,t,f,t)$

## 2.2.6 Modified Condition/Decision Coverage (MCDC)

"Every point of entry and exit in the program has been invoked at least once, every condition in a decision in the program has taken on all possible outcomes at least once, and each condition has been shown to independently affect the decision's outcome". MCDC pair for a condition is one that changes the output on varying the input from "f" to "t" while keeping the other conditions fixed. At least one pair for each condition is required to form the test suite. A condition is the occurrence of a variable in the Boolean expression.

The MC/DC coverage became popular after it was adopted as standard a standard requirement for airborne software. Chilenski and Miller have described applicability of MC/DC coverage to software testing Kapoor and Bowen has reported variations in the fault detection effectiveness of decision coverage (DC), full predicate coverage (FPC) and MC/DC coverage. They found that while average effectiveness of DC and FPC criteria decreases with the increase in the number of conditions in the program under test, it remains constant for MC/DC.

**Algorithm for Modified Condition/Decision Coverage (MC/DC)**
A test set T for program P
1. Cover each block in P
2. Each simple condition in P has taken both true and false values
3. Each decision in P has taken all possible outcomes
4. Each simple condition within a compound condition C in P independently effect the outcome of C.

**Example:**
Let $E=(ab) + c$
Set of test cases generated using MC/DC (t,f,t,t)(t,f,f,f)(t,t,f,t)(f,t,f,t)

### 2.2.7 MUMCUT strategy

MUMCUT strategy [10, 15, 35] integrates the Multiple Unique True Point (MUTP), Multiple Near False Point (MNFP) and Corresponding Unique True Point and Near False Point Pair (CUTPNFP). Boolean specifications need to be in irredundant disjunctive normal form.
MUTP strategy: Select test points in Unique True Point UTP(i) such that every truth value of every missing variable is covered.
MNFP strategy: Select test points in Near False Point NFP (i,j) such that every truth value of every missing variable is covered.
CUTPNFP strategy: Select a unique true point in UTP(i) and a near false point in NFP(i,j) such that the two points differ only at the $j^{th}$ literal of the $i^{th}$ term

**Example:**
Let $E=ab+cd$
Set of test cases generated using MUMCUT
By applying MUTP strategy{(t,t,f,t)(t,t,t,f)(f,t,t,t)(t,f,t,t)
By applying MNFP strategy{(f,t,f,t)(f,t,t,f)(t,f,f,t)(t,f,t,f)
By applying CUTPNFP strategy {(t,t,f,t)(f,t,f,t)(t,f,f,t)(f,t,t,t)(f,t,f,t)(f,,t,t,f)}

### 2.3 Fault based approach

The effectiveness of above mentioned strategies is mostly assessed in terms of their ability in identifying mutations. In this paper classical fault classes have been used. Fault based analysis of Boolean specification and its software implementation has been explored both empirically [7, 21, 24, 36] and formally [9, 26, 28].
The various kinds of faults which can effect any Boolean expression are classified into the following categories.

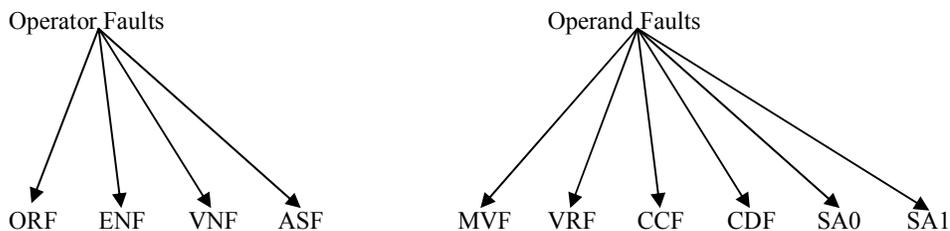

Figure 2: Classical 10 fault classes in to two categories, operator and operand faults.

### 2.3.1 Faults Categories

Operator Faults
- Operator Reference Fault (ORF): In this class of fault, a binary logical operator '.' is replaced by '+' or vice versa.

- Expression Negation Fault (ENF): A sub-expression in the statement is replaced by its negation (¬).
- Variable Negation Fault (VNF): An atomic Boolean literal is replaced by its negation (¬).
- Associative Shift Fault (ASF): This fault occurs when an association among conditions is incorrectly implemented due to misunderstanding about operator evaluation properties.
    - Parenthesis omission fault (POF): A pair of parentheses has been incorrectly omitted from the Boolean expression.
    - Parenthesis insertion fault (PIF): A pair of parentheses has been incorrectly inserted from the Boolean expression

Operand Faults
- Missing Variable Fault (MVF): A condition in the expression is missing with respect to original expression.
- Variable Reference Fault (VRF): A condition is replaced by another input which exists in the statement.
- Clause Conjunction Fault (CCF): A condition a in expression is replaced by *a.b,* where *b* is a variable in the expression.
- Clause Disjunction Fault (CDF): A condition *a* in expression is replaced with *a+b,* where b is a variable in the expression.
- Stuck at 0: A condition *a* is replaced with 0 in the function.
- Stuck at 1: A condition *a* is replaced with 1 in the function.

| Fault Type | Mutant example |
| --- | --- |
| ORF | (ab)(¬a+c) |
| ENF | ¬ (a+b)(¬a+c) |
| VNF | (¬a+b)(¬a+c) |
| ASF | (a+(b¬a)+c) |
| MVF | b(¬a+c) |
| VRF | (a+a)(¬a+c) |
| CCF | (a¬a+b)(¬a+c) |
| CDF | (a+c+b)(¬a+c) |
| SA0 (for a=0) | 0 |
| SA1 (for a=0) | bc |

Table2: Faults with their explanation for Boolean expression (a+b) (¬a+c)

**2.3.2 Fault Hierarchy**

A hierarchy among fault classes helps in generating tests; if test suite detects fault classes at the top of hierarchy, then all other faults in hierarchy will be detected by the same test suite.

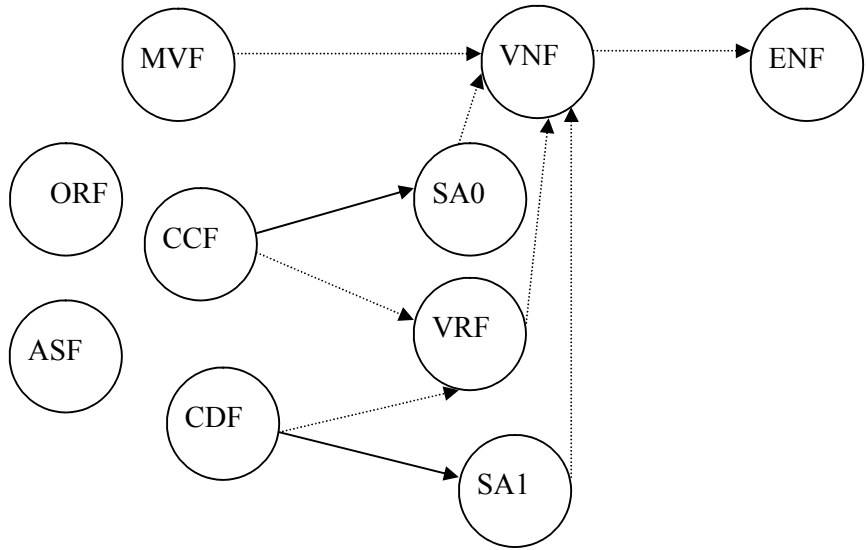

Figure 3: Hierarchy among fault classes: arrows among fault classes shows subsumption relations, dotted arrows represent subsumption relations which were initially established [7] but were later proved not to hold [33]

## 3 COMPARING BOOLEAN SPECIFICATION TESTING STRATEGIES

This section contains comparison of Boolean specification testing strategies with respect to fault detection ability and size of generated test suite.

### 3.1 Fault classes

Empirical evaluation of testing technique for Boolean specifications were studied in [8, 9, 11, 13] it has explored the relationship between various fault types. It is shown that ENF are weakest faults in the sense that any technique which catches stronger faults are likely to find ENF's.[28] improved the results, however the results are applicable only for associated faulty decisions.

### 3.2 Size of generated test suite

As exhaustive testing is not feasible, various techniques result in minimization of test suite, various papers have identified the impact of minimization of test set size on their fault detection effectiveness.

## 4 CONCLUSIONS

Much of the published research in fault class analysis was based on empirical evidences, an empirical evaluation of the BOR, Elmendorf's Strategy, MCDC using fault based approach has been performed [11]. Boolean expressions from literature ranging from 3 variables to 12 variables used for assessing the, performance and effectiveness of the various testing techniques based on mutation analysis. Mutated expressions were generated from the given Boolean expressions by making syntactic change based on particular type of fault. The results were in favour of Elmendorf Method for detection of all fault classes, but the size of test suite is large. BOR technique has been originally designed for the detection of missing/extra negation operators; therefore, it does not guarantee the detection of other faults. The other limitation of BOR technique is that it is suitable only for the singular expression and performs poorly in the cases where the expression has coupling effect. Performance of MCDC is much better than

BOR for all kinds of Faults. The size of the test suite is also comparable to BOR. Reported [8] that average effectiveness of MC/DC remains constant even with increase in number of conditions. MI and MUMCUT does not any restriction on number of variable and number of occurrences of the variables It has been shown [35] that MUMCUT detects all faults detected by MI and the test generated is a subset of test sets generated by MI and the size of test suit is much smaller. One approach to overcome the weakness of these is to combine these techniques.

**References**

1) B. Beizer. Software Testing Techniques. Van Nostrand Reinhold, Inc. New York, 2nd edition, 1990.

2) J. J. Chilenski. An Investigation of Three Forms of the Modified Condition Decision Coverage (MCDC) Criterion. Technical Report DOT/FAA/AR 01/18, U.S. Department of Transportation, Federal Aviation Administration, April 2001.

3) J.J. Chilenski and S. Miller. Applicability of Modified Condition/ Decision Coverage to Software Testing. Software Engineering Journal, 9(5):193 200, September 1994.

4) W. R. Elmendorf. Cause-Effect Graphs on Functional Testing, TR-00.2487, IBM Systems Development Division, Poughkeepsie, NY(1973)

5) P. G. Frankl and E. Weyuker. A Formal Analysis of the Fault- Detecting Ability of Testing Methods. IEEE Transactions on Software Engineering, 19(3):202–213, March 1993.

6) J. A. Jones and M. J. Harrold. Test-Suite Reduction and Prioritization for Modified Condition/Decision Coverage. In International Conference on Software Maintenance (ICSM), pages 92–101. IEEE, November 2001.

7) K. Kapoor and J. P. Bowen. Tconditions for fault classes in Boolean Specifications. ACM Transactions on software engineering and methodology; 2007

8) K. Kapoor and J. P. Bowen. Experimental Evaluation of the Variation in Effectiveness for DC, FPC, MC/DC Test Criteria. In International Symposium on Empirical Software Engineering (ISESE), pages 185–194, September 2003.

9) D.R. Kuhn. Fault classes and error detection capability of specification based testing technique. ACM Transactions on Software engineering and methodology, 8(4), 411-424, 1999

10) M.F Lau, Y. Liu and Y. T. Yu. An extended fault class hierarchy for specification based testing, ACM transactions on Software engineering and methodology, 14(3), 247-276, 2005

11) R.K. Singh, P. Chandra, Y. Singh. An Evaluation of Boolean Expression Testing Techniques, ACM transactions on Software engineering and methodology, 2006

12) M. Grindal, J. Offutt, S.F. Andler, Combination Testing Strategies: A survey, GMU Technical report ISE- TR-04-05, July 2004

13) A. Gargantini, G. Fraser, Generating Minimal Fault detecting Test suites for General Boolean Specifications, 2011

14) A. Paradkar, A New Solution to Test Generation for Boolean Expressions; 1995

15) T.Y. Chen, M.F lau, K.Y. Sim. C.A. Sun, On detecting faults for Boolean expressions, Software quality journal, 2008

16) K. A. Foster. ,Sensitive Test Data for Logic Expressions ". ACM SIGSOFT Software Eng.Notes, Vol. 9, No. 2, pages 120-26, April 1984

17) N. G. Leveson, M. P. E. Heimdahl, H. Hildreth,J. D. Reese. Requirements Specification for Process-control Systems". TR 92-106, Dept. of Inform. and Comp. Sci., Univ. of Cal., Irvine, Nov1992.

18) G. Myers. The Art of Software Testing. Wiley-Interscience, 1979.

19) A.F. Offutt. Investigations of the Software Testing Coupling Effect, ACM Transactions on Software Engineering and Methodology, Vol 1(1), pp 5-20, January 1992

20) A. Paradkar and K.C. Tai. Test-Generation for Boolean Expressions. In Sixth International Symposium on Software Reliability Engineering (ISSRE), pages 106–115, 1995.

21) A. Paradkar, K. C. Tai, and M. A. Vouk. Automatic Test-Generation for Predicates. IEEE Transactions on Reliability, 45(4):515–530, December 1996.

22) K.C. Tai. Theory of Fault-based Predicate Testing for Computer Programs. IEEE Transactions on Software Engineering, 22(8):552–562, August 1996.

23) K.C. Tai. Theory of Fault Based Predicate Testing for Computer Programs, IEEE Transactions of Software Engineering, vol 22, no 8, pp 552-562, 1996

24) K.C Tai. M.A Vouk., A. Paradkar., Lu P. , "Predicate Based Testing," IBM Systems Journal, Vol 33 (3), p 445, 1994

25) M. A. Vouk, K. C. Tai, and A. Paradkar. Empirical Studies of Predicate-based Software Testing. In 5th International Symposium on Software Reliability Engineering, pages 55–64. IEEE, 1994.

26) K. C. Tai. Predicate-Based Test Generation For Computer Programs ". Proceedings of International Conference on Software Engineering, pages 267-276, May 1993.

27) K. C. Tai, M. A. Vouk, Amit Paradkar, P.Lu. Evaluation of a Predicate-Based Software Testing Strategy". IBM Systems Journal, Vol.33, No.3, pages 445{457, October 1994.

28) T. Tsuchiya and T. Kikuno. On Fault Classes and Error Detection Capability of Specification-based Testing. ACM Transactions on Software Engineering and Methodology, 11(1):58–62, January 2002.

29) S. A. Vilkomir, K. Kapoor, and J. P. Bowen. Tolerance of Control- Flow Testing Criteria. In 27th International Computer Software and Applications Conference (COMPSAC), pages 182–187. IEEE Computer Society, November 2003.

30) Praveen Ranjan Srivastava, Prashad Patel, Siddharth hatrola; Cause Effect Graph to Decision Table generation, ACM SIGSOFT Software Eng.Notes, Vol. 34, No. 2,2009

31) A. P Matur, "Software testing", 1st edition, Pearson Publication, 2008

32) K. Nursimulu,R.L Probert, "Cause effect Graphing Analysis and validation requirements"

33) Z. Chen, and B. Xu. A revisit of fault class hierarchies in general Boolean specifications. ACM Transactions on Software Engineering and Methodology, 2010.

34) E. Weyuker, T. Gorodia, and A. Singh. Automatically Generating Test Data from a Boolean Specification. IEEE Transactions on Software Engineering, 20(5):353–363, May 1994.

35) T. Y. Chen & M.F. Lau. Test case selection strategies based on Boolean Specifications. Software Testing, Verification and Reliability, 2001

**Authors**

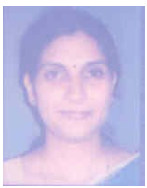
**Usha Badhera is** an active researcher in the filed of software testing, currently working as Assistant Professor in Department of Computer Science at Banasthali University (Rajasthan), India. She has done MCA from Rajasthan University and her PhD is in progress from Banasthali University (Rajasthan), India.

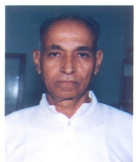
**Prof. G. N. Purohit** is a Professor in Department of Mathematics & Statistics at Banasthali University (Rajasthan). Before joining Banasthali University, he was Professor and Head of the Department of Mathematics, University of Rajasthan, Jaipur. He had been Chief-editor of a research journal and regular reviewer of many journals. His present interest is in O.R., Discrete Mathematics and Communication networks. He has published around 40 research papers in various journals.

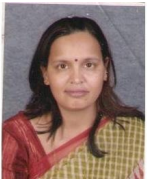 **S.Taruna is** an active researcher in the filed of communication and mobile network, currently working as Assistant Professor in Department of Computer Science at Banasthali University ( Rajasthan), India. She has done M.Sc from Rajasthan University and her PhD is in progress from Banasthali University(Rajasthan) , India.